\documentclass[reprint,amsmath,amssymb,aps]{revtex4-2}
\usepackage{amssymb}
\usepackage{graphicx}
\usepackage{dcolumn}
\usepackage{bm}
\usepackage{color}
\usepackage{subfigure}
\usepackage{xspace}
\usepackage{textpos}
\usepackage[english]{babel}
\usepackage[colorlinks,linkcolor=blue,anchorcolor=blue,citecolor=blue,urlcolor=blue]{hyperref}
\usepackage[all]{hypcap}

\newcommand\bes{BES\uppercase\expandafter{\romannumeral3}}
\newcommand\BEPC{BEPC\uppercase\expandafter{\romannumeral2}}

\newcommand{\errS}{4.3}

\newcommand{\Br}{7.3}
\newcommand{\Brerr}{0.8}
\newcommand{\Brstat}{1.8}

\def\arraystretch{0.8}

\begin{document}
\title{\boldmath Evidence of the \texorpdfstring{$h_c\to K_S^0 K^+\pi^-+c.c.$}{hc2KsKpi} decay}

\begin{abstract}
Based on $(2.712\pm0.014)\times10^9$ $\psi(3686)$ events collected by
the BESIII collaboration, evidence of the hadronic decay $h_c\to
K_S^0K^+\pi^-+c.c.$ is found with a significance of $4.3\sigma$ in the
$\psi(3686)\to\pi^0 h_c$ process. The branching fraction of $h_c\to
K_S^0 K^+\pi^- +c.c.$ is measured to be
$(\Br\pm\Brstat\pm\Brerr)\times10^{-4}$, where the first and second
uncertainties are statistical and systematic, respectively. Combining with
the exclusive decay width of $\eta_c\to K\bar{K}\pi$, our result
indicates inconsistencies with both pQCD and NRQCD predictions.
\end{abstract}

\date{\today}

\author{M.~Ablikim$^{1}$, M.~N.~Achasov$^{4,c}$, P.~Adlarson$^{75}$, O.~Afedulidis$^{3}$, X.~C.~Ai$^{80}$, R.~Aliberti$^{35}$, A.~Amoroso$^{74A,74C}$, Q.~An$^{71,58,a}$, Y.~Bai$^{57}$, O.~Bakina$^{36}$
% ...
\\
\vspace{0.2cm}
(BESIII Collaboration)\\
\vspace{0.2cm} {\it
$^{1}$ Institute of High Energy Physics, Beijing 100049, People's Republic of China\\
% ...
$^{3}$ Bochum  Ruhr-University, D-44780 Bochum, Germany\\
$^{4}$ Budker Institute of Nuclear Physics SB RAS (BINP), Novosibirsk 630090, Russia\\
% ...
$^{35}$ Johannes Gutenberg University of Mainz, Johann-Joachim-Becher-Weg 45, D-55099 Mainz, Germany\\
$^{36}$ Joint Institute for Nuclear Research, 141980 Dubna, Moscow region, Russia\\
% ...
$^{57}$ Southeast University, Nanjing 211100, People's Republic of China\\
$^{58}$ State Key Laboratory of Particle Detection and Electronics, Beijing 100049, Hefei 230026, People's Republic of China\\
% ...
$^{71}$ University of Science and Technology of China, Hefei 230026, People's Republic of China\\
% ...
$^{74}$ University of Turin and INFN, (A)University of Turin, I-10125, Turin, Italy; (B)University of Eastern Piedmont, I-15121, Alessandria, Italy; (C)INFN, I-10125, Turin, Italy\\
$^{75}$ Uppsala University, Box 516, SE-75120 Uppsala, Sweden\\
% ...
$^{80}$ Zhengzhou University, Zhengzhou 450001, People's Republic of China\\
\vspace{0.2cm}
$^{a}$ Deceased\\
% ...
$^{c}$ Also at the Novosibirsk State University, Novosibirsk, 630090, Russia\\
% ...
}
}



\maketitle

\section{INTRODUCTION}
\label{section:Introduction}
In the framework of the Standard Model, one fundamental assumption is that the strong interaction between quarks is mediated by colored gluons, as described by Quantum Chromodynamics (QCD). Since the derivative of QCD coupling strength ($\alpha_s$) with respect to the logarithm of the energy scale, i.e. the $\beta$  function, is negative~\cite{PhysRevLett.30.1343, PhysRevLett.30.1346}, $\alpha_s$ is divergent in the low-energy limit, bringing large uncertainties to the theoretical calculations in this non-perturbative region. Because of their large
mass scales and non-relativistic nature, charmonia are ideal probes to study and understand QCD from both perturbative and non-perturbative aspects.
Based on perturbative QCD (pQCD) and non-relativistic QCD (NRQCD), the branching fractions of several light hadron decay channels of the $P$-wave spin singlet charmonium $h_c(1P)$ were calculated~\cite{Kuang}. For the channel $K\bar{K}\pi$, the predictions are $(1.4\pm0.9)\%$ (pQCD) and $(5.5\pm3.3)\%$ (NRQCD), respectively. An experimental measurement on this channel is helpful for testing the validity of these approaches.

While a great deal of hadronic decay modes of the $S$-wave charmonia
including $J\slash\psi$ and $\psi(3686)$ have been discovered and
determined precisely, the knowledge of the hadronic decays of $h_c$ is
still sparse, since its direct production via $e^+e^-$ annihilation is
forbidden. Until now, only a few hadronic decay modes have been
observed~\cite{PhysRevD.99.072008, PhysRevD.102.112007} via the
$\psi(3686)\to \pi^0 h_c$ process, with a sum of branching fractions
less than $5\%$ \cite{PDG}. According to the experimental result on
the branching fraction of the prominent electromagnetic transition
$\mathcal{B}(h_c\to\gamma\eta_c)=(57.66^{+3.62}_{-3.50}\pm0.58)\%$~\cite{PhysRevD.106.072007},
the fraction of gluonic annihilation width $\mathcal{B}(h_c\to 3g)$ is
estimated to be about $42\%$, indicating that a significant fraction of
$h_c$ hadronic decay modes remain unknown. Furthermore, although the
decay $h_c\to K_S^0 K^+\pi^-$ has been predicted for a long
time~\cite{Kuang}, it has not yet been discovered, only an upper limit
of $6\times10^{-4}$ has been
determined~\cite{PhysRevD.102.112007,PDG}.

A search for the hadronic decay mode $h_c\to K_S^0 K^+\pi^-$ is
performed based on the $\psi(3686)\to\pi^0 h_c$ process from the data
sample with $2.712\pm0.014$ billion $\psi(3686)$
events~~\cite{psip2021} collected at center-of-mass energy
$\sqrt{s}=3.686$ GeV with the BESIII detector. Throughout this paper,
charged conjugation is always implied unless otherwise specified.

\section{BESIII DETECTOR AND MONTE CARLO
SIMULATION}
\label{sec:detector}

The BESIII detector~\cite{Ablikim:2009aa} records symmetric $e^+e^-$
collisions provided by the BEPCII storage ring~\cite{Yu:2016cof},
which operates with a center-of-mass energy range from $2.00$ to
$4.95$~GeV, with a peak luminosity of
$1.1\times10^{33}$~cm$^{-2}$s$^{-1}$ achieved at
$3.773$~GeV. \mbox{BESIII} has collected large data samples in this
energy region~\cite{Ablikim:2019hff,EcmsMea,EventFilter}. The
cylindrical core of the BESIII detector covers 93\% of the full solid
angle and consists of a helium-based multilayer drift chamber~(MDC), a
plastic scintillator time-of-flight system~(TOF), and a CsI(Tl)
electromagnetic calorimeter~(EMC), which are all enclosed in a
superconducting solenoidal magnet providing a 1.0~T magnetic
field. The solenoid is supported by an octagonal flux-return yoke with
resistive plate counter muon-identification modules interleaved with
steel. The charged-particle momentum resolution at $1~{\rm GeV}/c$ is
$0.5\%$, and the resolution of the rate of energy loss, ${\rm d}E/{\rm
  d}x$, is $6\%$ for electrons from Bhabha scattering. The EMC
measures photon energies with a resolution of $2.5\%$ ($5\%$) at
$1$~GeV in the barrel (end-cap) region. The time resolution in the TOF
barrel region is 68~ps, while that in the end-cap region was 110~ps.
The end cap TOF system was upgraded in 2015 using multigap resistive
plate chamber technology, providing a time resolution of 60~ps, which
benefits 80\% of the data used in this
analysis~\cite{etof1,etof2,etof3}.

Simulated data samples produced with a {\sc
  geant4}-based~\cite{geant4} Monte Carlo (MC), which includes the
geometric description of the BESIII detector and the detector
response, are used to determine detection efficiencies and to estimate
backgrounds. The simulation models the beam-energy spread and
initial-state radiation in the $e^+e^-$ annihilations with the
generator {\sc kkmc}~\cite{ref:kkmc}. The inclusive MC sample includes
the production of the $\psi(3686)$ resonance, the ISR production of
the $J/\psi$, and the continuum processes incorporated in {\sc
  kkmc}~\cite{ref:kkmc}. All particle decays are modeled with {\sc
  evtgen}~\cite{Lange:2001uf,Ping:2008zz} using branching fractions
either taken from the Particle Data Group~\cite{PDG}, when available,
or otherwise estimated with {\sc
  lundcharm}~\cite{YANGRui-Ling:61301,Chen:2000tv}. Final-state
radiation from charged final-state particles is incorporated using
{\sc photos}~\cite{RICHTERWAS1993163}.  Additionally, an exclusive MC
sample of the signal process $\psi(3686)\to\pi^0h_c,~h_c\to
  K_S^0K^+\pi^-$ is generated for the selection criteria optimization
and detection efficiency determination, where both $\psi(3686)\to\pi^0
h_c$ and $h_c\to K_S^0K^+\pi^-$ are simulated based on a uniform phase space
distribution.

\section{DATA ANALYSIS}
\label{sec:DATA_ANALYSIS}

\subsection{Event Selection}
\label{sec:Event_Selection}

To select $\pi^0K_S^0K^+\pi^-$ events, all the charged tracks are
required to be within a polar angle ($\theta$) range of
$\vert\!\cos\theta\vert<0.93$, where $\theta$ is measured with respect to the
$z$-axis, i.e., the symmetry axis of MDC. Exactly four charged tracks
are required after the polar angle selection, and the total net charge
is required to be zero.

To reconstruct $K_S^0$, all pairs of oppositely charged tracks
are subjected to a secondary vertex fit,
which constrains the two tracks to originate from a common vertex. The
pair with invariant mass closest to the $K_S^0$ known mass is
chosen, and the distance of the common vertex to the interaction point
normalized by its uncertainty, $L\slash\Delta L$, is required to be
greater than 2.

For the charged tracks not originating from $K_S^0$ decays, the distance of closest approach to the interaction point (IP)  must be less than 10\,cm along the $z$-axis, and less than 1\,cm in the transverse plane. For each track, particle identification (PID) chi-square values $\chi^2_{\rm PID}(\mathcal{P_\pm})$ under different particle hypotheses are computed utilizing the combined information of 
$dE/dx$ and TOF, where $\mathcal{P}\in\{K,\pi \}$ refers to the particle types and the subscript indicates the positive or negative charge of the particle.

Photon candidates are reconstructed from electromagnetic clusters produced in the crystals of the EMC. Clusters with deposited energy larger than 25 MeV in the barrel region ($\vert\!\cos\theta\vert<0.8$) or 50 MeV in the end cap region ($0.86<\vert\!\cos\theta\vert<0.92$) are selected as photon candidates. The EMC timing of each photon counted from the event start time is required to be within [0, 700] ns to suppress noise and energy deposits unrelated to the event. The opening angle between the photon and the nearest charged track is required to be
larger than $10^\circ$ to suppress the showers from charged tracks.

A pair of photons is accepted as a $\pi^0$ candidate if their invariant
mass falls into the range $(0.12,~0.15)~{\rm MeV}/c^2$. At least one
$\pi^0$ is required. All combinations of final state particles are
subjected to a five-constraint kinematic fit, with the constraints
provided by four-momentum conservation and the $\pi^0$ mass. PID of the
two tracks and the selection of the best $\pi^0$ among multiple
candidates are accomplished by minimizing $\chi^2=\chi^2_{5 {\rm
    C}}+\chi^2_{\rm PID}(\mathcal{P_+})+\chi^2_{\rm
  PID}(\mathcal{P_-})$, where $\chi^2_{5 {\rm C}}$ is the fit quality
of the 5C kinematic fit. The combination with the minimum $\chi^2$
and satisfying $\mathcal{P}_+\mathcal{P}_-=K^+\pi^-$ or $K^-\pi^+$ is accepted
as a signal candidate. Additionally, $\chi^2_{5 {\rm C}}<124$ is
required to further suppress the non-$\pi^0K_S^0K^+\pi^-$ background.

To suppress background with one more or one less photon in the final
state, a four-constraint kinematic fit with the constraint provided
by four-momentum conservation is performed, and the chi-square values
with one or three photons $\chi^2_{4 {\rm C}}(1\gamma)$ and $\chi^2_{4
  {\rm C}}(3\gamma)$ are required to be larger than that of the signal
candidate, where the test is done for all the photon
candidates. Background $\pi^0$ produced from $K^*(892)^+\to K^+\pi^0$
are found in the inclusive MC sample and removed by applying
$\vert M(K^+\pi^0)-m(K^*(892)^+)\vert>10~{\rm MeV}/c^2$, where $M(K^+\pi^0)$ is
the invariant mass of $K^+\pi^0$ and $m(K^*(892)^+)$ is the known
mass of $K^*(892)^+$~\cite{PDG}. All the selection criteria have been
optimized by maximizing the figure of merit defined in
Ref.~\cite{Punzi:2003bu}, $ \epsilon\slash( a/2+\sqrt{B})$, where
$a=5$ is the desired one-tailed significance, $\epsilon$ is the
selection efficiency and $B$ is the expected number of background
events falling in the signal range from MC simulation. Here, the
signal range is defined as $\vert M(K_S^0K^+\pi^-)-m(h_c)\vert<6~{\rm
  MeV}/c^2$, where $M(K_SK^+\pi^-)$ is the invariant mass of
$K_SK^+\pi^-$, $m(h_c)$ is the $h_c$ known mass~\cite{PDG}, and the
interval is determined to be about $\pm 2 \sigma$ with respect to the
resolution.

\subsection{Fit to Data}
\label{section:Fit}

\begin{figure}[htbp]\centering
\includegraphics[height=6cm]{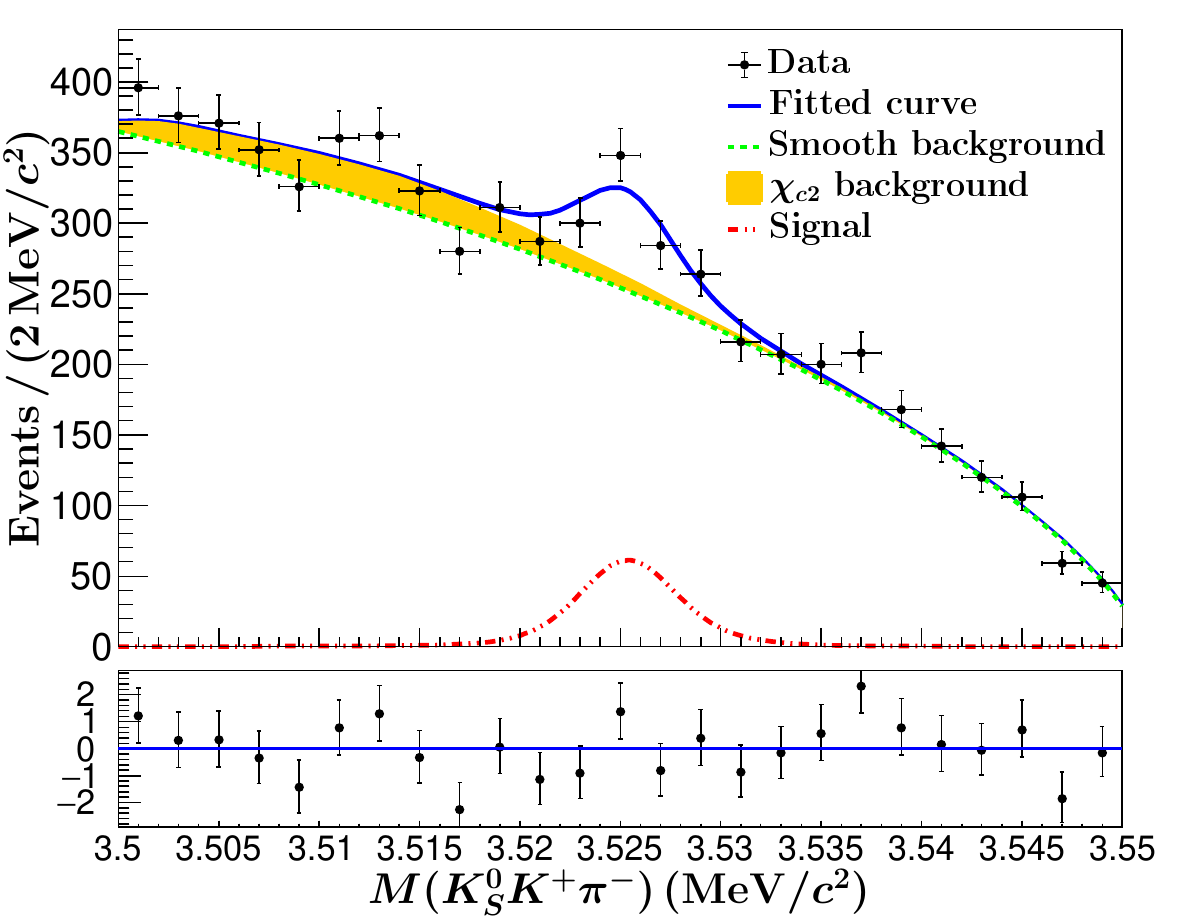}
\caption{Fit to the $M(K_S^0K^+\pi^-)$ distribution. The black dots with error bars are data, the blue solid curve is the total fit result, the red dashed-dotted curve is the $h_c$ signal, the green dashed  curve represents the smooth background, and the orange-filled shape represents the $\chi_{c2}$ peaking background.}\label{fig:fit}
\end{figure}

To determine the number of signal events $N$, an unbinned
maximum-likelihood fit is performed to the $M(K_SK^+\pi^-)$
distribution, as shown in Fig.~\ref{fig:fit}. In the fit, the $h_c$
signal is modeled by the shape obtained from the signal MC sample
convolved with a Gaussian function which accounts for the difference
in the mass resolution between MC simulation and data. The central
value and standard deviation of the Gaussian function are fixed to
those obtained by studying the control sample of
$\psi(3686)\to\gamma\chi_{c1,c2},~ \chi_{c1,c2}\to K_S^0K^+\pi^-$.  A peaking
background component from $\psi(3686)\to \gamma\chi_{c2},~\chi_{c2}\to
  K_S^0K^+\pi^-$ plus a fake photon, which forms a peak at the left
side of the $h_c$ signal, is included in the fit. The number of
peaking background events is fixed based on the measured branching
fraction~\cite{PDG}. Background from other channels distributes
smoothly and is described by an ARGUS function~\cite{ALBRECHT1990278},
with the threshold parameter fixed to the kinematic threshold 3.551
${\rm GeV}/c^2$.  The fit gives $N=205\pm50$ with a statistical
significance $4.6\sigma$, determined by the likelihood ratio of the
fit with and without the $h_c$ signal function~\cite{Wilks}.

The branching fraction is calculated with
\begin{equation}\label{eq:BF}\begin{split}
\mathcal{B}(h_c\to K_S^0K^+\pi^-)\times\mathcal{B}(\psi(3686))=
\frac{N}{\epsilon N_{\psi(3686)} \mathcal{B}(K_S^0) \mathcal{B}(\pi^0)},
\end{split}
\end{equation}
where $\epsilon=20.8\%$ is the selection efficiency, $N_{\psi(3686)}$
is the number of $\psi(3686)$ events~\cite{psip2021},
$\mathcal{B}(K_S^0)$ and $\mathcal{B}(\pi^0)$ are the branching
fractions of $K_S^0\to\pi^+\pi^-$ and $\pi^0\to\gamma\gamma$,
respectively, from the particle data group~\cite{PDG}, and
$\mathcal{B}(\psi(3686))$ is the branching fraction of
$\psi(3686)\to\pi^0h_c$ from Ref.~\cite{PhysRevD.106.072007}. The
resulting values are $\mathcal{B}(h_c\to
K_S^0K^+\pi^-)=(7.3\pm1.8)\times10^{-4}$ and $\mathcal{B}(h_c\to
K_S^0K^+\pi^-)\times\mathcal{B}(\psi(3686))=(5.3\pm1.3)\times10^{-7}$
with statistical uncertainties only. 

\section{Systematic uncertainties}
\label{section:Systematic_uncertainties}

The sources of systematic uncertainties in the branching fraction measurement are tracking and PID efficiencies, reconstruction of photons, $\pi^0$ and $K_S^0$, $K^+\pi^0$ mass ranges, mass resolution of the signal shape, MC model, input values and fit method. All sources of systematic uncertainties, which are summarized in Table~\ref{tab:systematics}, are treated as independent and summed in quadrature. The details of their estimation are discussed in the following.
\begin{table}[htbp]
\renewcommand\arraystretch{1.5}
\caption{\label{tab:systematics}Summary of systematic uncertainties.}
\centering
\begin{tabular}{l|c}
\hline\hline
Source&Relative uncertainty\\
\hline
Tracking &2.0\% \\
PID&2.1\%\\
Photon reconstruction&2.0\%\\
$K_S^0$ reconstruction &1.0\%\\
$\pi^0$ reconstruction &0.7\%\\
5C Kinematic fit&0.9\%\\
MC model &4.9\%\\
$M(K^+\pi^0)$ mass window &$-$\\
Mass resolution &4.4\%\\
Fit method &3.4\%\\
Number of $\psi(3686)$ &0.5\% \\
$\mathcal{B}(\psi(3686))$&7.3\%\\\hline
Sum &11.1\% \\ \hline\hline
\end{tabular}
\end{table}

\paragraph*{Tracking efficiencies.} 
The uncertainties of tracking efficiencies of the kaon and pion from
the primary vertex are $1.0\%$ per track~\cite{PhysRevD.83.112005} and
added linearly.

\paragraph*{PID efficiency.}
The uncertainty of PID is studied with a $\psi(2S)\to
K_S^0K^\pm\pi^\mp\pi^0$ control sample. The sample is selected by
reconstructing one $K_S^0$ with a stricter requirement $L/\Delta L>20$
and one $\pi^0$ with the number of photons restricted to be two. The
signal region is $0.12<M(\gamma\gamma)<0.14~{\rm GeV}/c^{2}$
and $487.6<M(K_S^0)<507.6~{\rm MeV}/c^{2}$, and the PID efficiency is
defined as $N_{\rm PID}/N_{\rm all}$, where $N_{\rm PID}$ and $N_{\rm
  all}$ are the number of events in the signal region with and without
PID. The difference between the efficiencies of MC and data is take as
the systematic uncertainty.

\paragraph*{Photon reconstruction.}
The uncertainty of photon reconstruction, which is $1.0$\% for each
photon, was determined with the control sample of $J\slash\psi\to
\rho^0\pi^0,~\rho^0\to\pi^+\pi^-,~\pi^0\to\gamma\gamma$~\cite{PhysRevD.81.052005}.

\paragraph*{$K_S^0$ reconstruction.} 
The difference of $K_S^0$ reconstruction between MC simulation and
data was studied based on the control samples of $J\slash\psi\to
  K^*(892)^\pm K^\mp,~K^*(892)^\pm \to K_S^0\pi^\pm$ and $J\slash\psi\to\phi
  K_S^0K^\pm\pi^\mp$~\cite{PhysRevD.92.112008}, and its related
uncertainty is $1.0$\%.

\paragraph*{$\pi^0$ reconstruction.}
The uncertainty of $\pi^0$ reconstruction is studied using
$\psi(3686)\to K_S^0K^+\pi^-\pi^0$ reconstructed with and without
a $\pi^0$ mass constraint $(0.12, 0.15)~{\rm
  GeV}/c^2$ and with the events in $h_c$ signal range excluded. The
difference of $\pi^0$ reconstruction efficiency between data and MC
simulation gives a relative uncertainty of $0.7\%$.

\paragraph*{Kinematic fit.} 
A correction of the helix parameters of the charged tracks in the MC
samples is applied to improve the consistency between MC simulation
and data. The uncertainty is estimated as half of the difference of
efficiency with and without the correction, as suggested in
Ref.~\cite{PhysRevD.87.012002}, resulting in an uncertainty of 0.9\%.

\paragraph*{MC model.} Because of the limited knowledge of intermediate states in the $h_c$ decays, alternative MC samples are generated including the intermediate state $K^*(892)$ with the fractions of neutral and charged channels constrained by isospin symmetry. The result is compared with that of the nominal phase space result, and the difference is treated as the systematic uncertainty.

\paragraph*{Mass resolution.} 
An alternative control sample, $\psi(3686)\to \pi^0 J\slash\psi,
J\slash\psi\to K_S^0K^\pm\pi^{\mp}$ is used to determine the
parameters of the mass resolution function, and the fit is
repeated with the alternative mass resolution function. The difference
of the fit yields to the nominal value gives a systematic uncertainty
of $4.4$\%.

\paragraph*{$K^+\pi^0$ mass window.} To study the potential uncertainty arising from $M(K^+\pi^0)$  mass window, a test introduced in Ref.~\cite{barlow2002systematic} is performed by varying the range of the mass window around its nominal width. The test statistic $\xi$ is defined as $\xi=\frac{\vert\mathcal{B}-\mathcal{B}^{\prime}\vert}{ \sqrt{\vert\sigma^2-\sigma^{\prime2}\vert}}$ where $\mathcal{B}$ ($\sigma$) and $\mathcal{B}^{\prime}$ ($\sigma^{\prime}$) are the resulting branching fractions (uncertainties) of the nominal result and that from varied mass windows, respectively. If the values of $\xi$ show a trending behavior and exceed two, a systematic uncertainty is assigned. We find the values of $\xi$ of the branching fraction are always less than two and show no trending behavior around the nominal mass window, therefore the related uncertainty is negligible.

\paragraph*{Fit method.}
The same test as described in the last paragraph is performed with
respect to the fitting ranges, and the related systematic effect is
proved to be negligible. For the background modeling, the shape for
the smooth background is changed from an ARGUS function to a second-order Chebychev polynomial, and for the peaking background, the
normalization scale is varied according to the uncertainty of the
world average value of the branching fraction
$\psi(3686)\to\gamma\chi_{c2},~\chi_{c2}\to K_S^0K^+\pi^-$~\cite{PDG}. The
corresponding change of the result is taken as the systematic
uncertainty.

\paragraph*{Input values.} 
The number of $\psi(3686)$ events is quoted from
Ref.~\cite{psip2021}, and the related uncertainty is taken into
account. For the input branching fractions, the uncertainties
associated with $K_S^0\to\pi^+\pi^-$ and $\pi^0\to \gamma\gamma$ are
negligible and the uncertainty of $\psi(3686)\to \pi^0h_c$ is assigned
according to Ref.~\cite{PhysRevD.106.072007}.

The signal significance is re-estimated after considering alternative
$M(K^+\pi^0)$ mass windows, mass resolution parameters, fitting ranges
and background shape. Among these tests, the lowest significance,
$4.3\sigma$, is taken as the final signal significance.
\section{SUMMARY AND DISCUSSION}
\label{section:Summary}

Based on $2.712\pm0.014$ billion $\psi(3686)$ events collected at
$\sqrt s=3.686$ GeV with the BESIII detector, evidence of the
hadronic decay $h_c\to K_S^0 K^+\pi^-$ is found with a significance of
$\errS\sigma$. The product branching fraction $\mathcal{B}(\psi(3686)\to\pi^0h_c)\times\mathcal{B}(h_c\to
K_S^0K^+\pi^-+c.c.)$ is
$(5.3\pm1.3\pm0.4)\times 10^{-7}$, and the branching fraction
$\mathcal{B}(h_c\to K_S^0K^+\pi^-+c.c.)$ is determined to be
$(\Br\pm\Brstat\pm\Brerr)\times10^{-4}$, where the first and second
uncertainties are statistical and systematic, respectively. Compared to the previous study by
BESIII~\cite{PhysRevD.102.112007}, in which an upper limit of
$\mathcal{B}(\psi(3686)\to\pi^0h_c)\times\mathcal{B}(h_c\to
K_S^0K^+\pi^-+c.c.)<4.8\times10^{-7},~CL=90\%$ is
obtained based on $4.48\times10^8$ $\psi(3686)$ events, our result is
consistent.

Assuming that isospin symmetry holds in the decay of $h_c\to
K\bar{K}\pi$, the ratios between isospin multiplets are
$\mathcal{B}(h_c\to K^0K^-\pi^+):\mathcal{B}(h_c\to \bar{K}^0K^+\pi^-):\mathcal{B}(h_c\to K^+K^-\pi^0):\mathcal{B}(h_c\to K^0\bar{K}^0\pi^0)=2:2:1:1$. Combining
with ${\mathcal{B}(K^0\to K_S^0)=0.5}$, we get the branching fraction
$\mathcal{B}(h_c\to K\bar{K}\pi)= 3 \times \mathcal{B}(h_c\to
K_S^0K^+\pi^-+c.c.)= (0.22\pm0.06)\%$, which is listed in
Table~\ref{tab:PDGKKpipi}. Our result is consistent with the
predictions based on pQCD and NRQCD, within $1.3$ and $1.6$ times the
uncertainties, respectively.

\begin{table}[htbp]
\renewcommand\arraystretch{1.5}
\caption{\label{tab:PDGKKpipi}The comparison between the theoretical predictions and the measurement of this work. Both statistical and systematic  uncertainties are included.}
\centering \begin{tabular}{c|c|c}
\hline\hline
Item&Value ($\%$)&Source\\ \hline
$\mathcal{B}(h_c\to K\bar{K}\pi)$&$1.4\pm0.9$&pQCD~\cite{Kuang}\\
$\mathcal{B}(h_c\to K\bar{K}\pi)$&$5.5\pm3.3$&NRQCD~\cite{Kuang}\\
$\mathcal{B}(h_c\to K\bar{K}\pi)$&$0.22\pm0.06$&This work\\
\hline
$\Gamma(h_c\to 3g)\slash\Gamma(\eta_c\to 2g)$&$1.0\pm0.1$&pQCD~\cite{Kuang}\\
$\Gamma(h_c\to 3g)\slash\Gamma(\eta_c\to 2g)$&$8.3\pm1.8$&NRQCD~\cite{Kuang}\\
$\Gamma(h_c\to 3g)\slash\Gamma(\eta_c\to 2g)$&$0.93\pm0.54$&Refs.~\cite{PDG,Wang:2021dxw}\\
$\Gamma(h_c\to K\bar{K}\pi)\slash\Gamma(\eta_c\to K\bar{K}\pi)$&$0.069\pm0.062$&This work\\
\hline\hline
\end{tabular}
\end{table}

The following relation~\cite{PhysRevD.46.2918,PhysRevD.37.1210} is
used for calculating the exclusive branching fractions in
Ref.~\cite{Kuang},
\begin{equation}\label{eq:rt}
\frac{\Gamma(h_c\to h)}{\Gamma(\eta_c\to h)}\approx\frac{\Gamma(h_c\to 3g)}{\Gamma(\eta_c\to 2g)},
\end{equation}
where $h$ is any exclusive hadronic channel and $\Gamma(h_c\to 3g)$
and $\Gamma(\eta_c\to 2g)$ are the inclusive gluonic annihilation
widths of $h_c$ and $\eta_c$, respectively. Taking
$\Gamma(h_c\to3g)\approx\Gamma(h_c)\times(1-\mathcal{B}(h_c\to\gamma\eta_c))\approx0.30\pm0.17$
MeV by neglecting all other radiative channels of $h_c$, and
$\Gamma(\eta_c\to2g)\approx\Gamma(\eta_c)\approx32.0\pm0.7$ MeV by
assuming a negligible radiative partial width of $\eta_c$, the ratio
in the right-hand side of Eq.~(\ref{eq:rt}) is $\frac{\Gamma(h_c\to
  3g)}{\Gamma(\eta_c\to 2g)}\approx(0.93\pm0.54)\%$, which is in good
agreement with the pQCD prediction $(1.0\pm0.1)\%$, but much smaller
than the NRQCD prediction of $(8.3\pm1.8)\%$~\cite{Kuang}. Using the
result from this work and the result from a global fit of $\eta_c$
branching fractions~\cite{Wang:2021dxw} as inputs, we determine the
ratio of partial widths $\frac{\Gamma(h_c\to
  K\bar{K}\pi)}{\Gamma(\eta_c\to K\bar{K}\pi)}=(0.069\pm0.062)\%$,
where the central value differs from the right-hand side of
Eq.~(\ref{eq:rt}) by one order of magnitude.

As a consequence, the test with Eq.~(\ref{eq:rt}) using our result for
$\Gamma(h_c\to K\bar{K}\pi)\slash\Gamma(\eta_c\to K\bar{K}\pi)$, which
is more sensitive than that of 
$\mathcal{B}(h_c\to K\bar{K}\pi)$, indicates
inconsistencies with both pQCD and NRQCD predictions.  Given the
predictions of the theoretical models have large uncertainties and
only leading-order formulas for both pQCD and NRQCD are implemented,
to resolve the gap between theoretical predictions and experimental
measurements, refined predictions are required with improved precision
and involving high-order effects such as corrections of
renormalization scale~\cite{CPL051202} or relativistic
effects~\cite{PhysRevD.88.034002}. On the other hand, the precision of
our measurement is still strongly limited by statistical uncertainty,
preventing us from drawing a solid conclusion. 

\section*{ACKNOWLEDGMENT}
The BESIII Collaboration thanks the staff of BEPCII and the IHEP computing center for their strong support. This work is supported in part by National Key R\&D Program of China under Contracts Nos. 2020YFA0406300, 2020YFA0406400, 2023YFA1606000; National Natural Science Foundation of China (NSFC) under Contracts Nos. 11635010, 11735014, 11835012, 11935015, 11935016, 11935018, 11961141012, 12025502, 12035009, 12035013, 12061131003, 12192260, 12192261, 12192262, 12192263, 12192264, 12192265, 12221005, 12225509, 12235017; the Chinese Academy of Sciences (CAS) Large-Scale Scientific Facility Program; the CAS Center for Excellence in Particle Physics (CCEPP); Joint Large-Scale Scientific Facility Funds of the NSFC and CAS under Contract No. U1832207; CAS Key Research Program of Frontier Sciences under Contracts Nos. QYZDJ-SSW-SLH003, QYZDJ-SSW-SLH040; 100 Talents Program of CAS; The Institute of Nuclear and Particle Physics (INPAC) and Shanghai Key Laboratory for Particle Physics and Cosmology; European Union's Horizon 2020 research and innovation programme under Marie Sklodowska-Curie grant agreement under Contract No. 894790; German Research Foundation DFG under Contracts Nos. 455635585, Collaborative Research Center CRC 1044, FOR5327, GRK 2149; Istituto Nazionale di Fisica Nucleare, Italy; Ministry of Development of Turkey under Contract No. DPT2006K-120470; National Research Foundation of Korea under Contract No. NRF-2022R1A2C1092335; National Science and Technology fund of Mongolia; National Science Research and Innovation Fund (NSRF) via the Program Management Unit for Human Resources \& Institutional Development, Research and Innovation of Thailand under Contract No. B16F640076; Polish National Science Centre under Contract No. 2019/35/O/ST2/02907; The Swedish Research Council; U. S. Department of Energy under Contract No. DE-FG02-05ER41374.

\bibliographystyle{apsrev4-2-2}
\bibliography{hcref.bib}

\end{document}